\documentclass{article}

\usepackage{amsmath, amssymb, amsthm}

\usepackage{color}
\usepackage{graphicx,float}% Include figure files
\usepackage{lipsum}
\usepackage{amsfonts}
\usepackage{graphicx}
\usepackage{epstopdf}
\usepackage{algorithmic}
\usepackage{url}

\theoremstyle{plain}
\newtheorem{theorem}{Theorem}
\newtheorem{rem}{Remark}
\newtheorem{Def}{Definition}

\newcommand{\be}{\begin{equation}}
\newcommand{\ee}{\end{equation}}
\newcommand{\xv}{{\bf x}}
\newcommand{\nv}{{\bf n}}
\newcommand{\vv}{{\bf v}}
\newcommand{\bef}{{\bf f}}
\newcommand{\ev}{{\bf e}}

\newcommand{\qv}{{\bf q}}

\newcommand{\de}{{\rm d}}
\newcommand{\hf}{{\bf F}}
\newcommand{\Fv}{{\bf F}}

\newcommand{\Mv}{{\bf M}}
\def\yv{{\bf y}}
\def\bv{{\bf b}}
\def\gv{{\bf g}}
\def\Lv{{\bf L}}
\def\bn{{\bf n}}

\def\phij{\varphi_{j}}
\def\phijd{\dot{\varphi}_{j}}
\def\alphaj{\alpha_j}

\numberwithin{equation}{section}

\def\aj{\alpha_{j}}
\def\ap{\alpha_{+1}}
\def\am{\alpha_{-1}}
\def\bap{\bar\alpha_{+1}}
\def\bam{\bar\alpha_{-1}}
\def\fj{\bef_j}
\def\vj{\vv_j}
\def\ej{\ev_j}
\def\nj{\bn_j}
\def\xj{\xv_j}

 \numberwithin{equation}{section}

\begin{document}

\title{Controllability and Displacement Analysis of a Three-Link Elastic Microswimmer: A Geometric Control Approach}

% Force line breaks with \\
\author{R. Attanasi \thanks{Dipartimento di Matematica e Fisica "Ennio De Giorgi", Università del Salento, Lecce, Italy} \and M. Zoppello \thanks{Dipartimento di Scienze Matematiche “Giuseppe Luigi Lagrange”, Politecnico di Torino, 10129 Torino, Italy} \and G. Napoli \thanks{ipartimento di Matematica e Applicazioni “Renato Caccioppoli”, Università degli Studi di Napoli “Federico II”, 80125 Napoli, Italy}}
\maketitle

\begin{abstract}
This study investigates the dynamics and controllability of a Purcell three-link microswimmer equipped with passive elastic torsional coils at its joints. By controlling the spontaneous curvature, we analyse the swimmer’s motion using both linear and weakly nonlinear approaches. Linear analysis reveals steady harmonic solutions for small-amplitude controls but does not predict any net displacement, whereas weakly nonlinear analysis predicts translation along the orientation of the central link. Using geometric control theory, we prove that the system is small time locally controllable near equilibrium and derive displacement estimates for periodic piecewise constant controls, which are validated through numerical simulations. These findings indicate that oscillatory controls can enable motion in all directions near equilibrium. This work offers foundational insights into the controllability of elastic microswimmers, paving the way for advanced motion planning and control strategies.
\end{abstract}

\maketitle

\section*{Introduction}
 Microswimmers, especially those inspired by biological systems, have attracted significant attention in recent years because of their potential applications in targeted drug delivery, microscale manipulation, and diagnostic procedures \cite{peyer2013bio, elgeti2015physics}.
These miniature robotic swimmers are designed to rely on sophisticated control strategies to navigate viscous environments where traditional propulsion mechanisms are either inefficient or infeasible, as exemplified by the Scallop Theorem \cite{Purcell1977}.
The study of their motion and controllability is crucial for advancing micro-robotic technologies, enhancing their practical applications in medical and industrial fields \cite{ 6600934, 8206004,bagagiolo2017swimming,PhysRevE.101.042604}.

A fundamental principle governing the motion of microswimmers is their operation in the very low Reynolds number regime, where inertial effects are negligible. Several basic models of microswimmers have been developed, starting with the three-link swimmer proposed by Purcell \cite{Purcell1977}, followed by the $N$-link swimmer \cite{alouges2015can, ZoppelloMorandotti22}, and systems of coupled microswimmers, as explored in \cite{zoppello2022controlling,attanasi2024purcell}.
With regard to the Purcell three-link swimmer, the work of \cite{passov2012dynamics} investigated the dynamics and stability of periodic solutions for the joint angles, assuming controlled inputs in the form of periodic torques applied at the joints. However, in many practical scenarios, the shape change of microswimmers occurs passively and is also affected by the elastic response of the swimmer's structure. Some models have studied the continuous bending deformation of elastic flagella with distributed actuation \cite{Lauga2007}, while others introduce a coarse-graining formulation in the asymptotic limit, where small rod-like elements are joined by torsional springs \cite{moreau2018asymptotic}.

In this paper, we investigate an enhanced version of the Purcell swimmer equipped with torsional springs at the joints, which we will subsequently refer to as an {\it elastic microswimmer}. These springs promote specific alignment angles between adjacent links, meaning that, in the absence of external forces, they induce a form of swimmer equilibrium dictated by the springs' resting angles. Consequently, the swimmer shows its own {\it spontaneous shape} (or {\it equilibrium shape}), which we assume can be controlled. The combined action of the torsional springs, which tend to restore the swimmer's form to its spontaneous shape, and the fluid's response to the rate of shape change enables movement.

We would like to emphasise the distinction between our approach and previous studies in the literature: unlike previous studies, we do not presume the swimmer's shape at each instant, that is, by prescribing the joint angles (as in \cite{passov2012dynamics}), nor do we control the torque at the joints (as in \cite{PhysRevE.110.014207}). Our control assumptions are less stringent, as we only specify the desired shape of the swimmer in the absence of external forces.

In this study, we investigate the dynamics of a Purcell three-link elastic microswimmer, focusing on the effects of assigning the springs' resting angles as control inputs. 
First of all our analysis involves the linearisation of the equations of motion around the swimmer's straight configuration. This requires that the controlled resting angles remain small throughout the motion.
For small-amplitude sinusoidal control gaits, we explicitly integrate the linearised equations and demonstrate that the system converges to a steady periodic trajectory. Then, only using a weakly nonlinear analysis, we are able to show that small sinusoidal controls result in a net translation in the direction of the central link.

Further analysis employs methods from geometric control theory: we rigorously explore the system's controllability properties. Our focus is on the small-time local controllability of the system, an essential aspect of manoeuvring at microscopic scales.
%The study employs both linear and nonlinear techniques to analyse controllability, acknowledging limitations in linearisation and the need for more sophisticated methods in specific contexts. 
We show that the elastic microswimmer exhibits small-time local controllability, enabling accurate steering between fixed configurations near equilibrium. Furthermore, we provide estimates of the microswimmer's displacement under both out-of-equilibrium and equilibrium conditions, using piecewise constant control inputs.

To validate our theoretical findings, we present numerical simulations that highlight the similarity in behaviour of the system under small sinusoidal controls and periodic piecewise constant controls. Our results suggest that predictions of microswimmer displacement using piecewise constant controls are also applicable to systems with continuous oscillating controls, particularly over long timescales and short control periods. 

The paper is organised as follows: in Section \ref{sec: Model}, we define the mathematical model and formulate the equations of motion governing the three-link elastic microswimmer. In Section \ref{sec:2}, we analyse the solution in the limit of small-amplitude oscillating control inputs. The solution obtained from weakly nonlinear analysis is compared with the numerical solution of the full nonlinear equation of motion. In Section \ref{sec:Controllability}, we first prove the local controllability of the microswimmer's motion over short times. We then use the method of Lie brackets to estimate the swimmer's displacement after one swimming cycle under partial constant controls. Finally, we compare the numerical solutions of the complete nonlinear problem with piecewise periodic controls and the solutions obtained by repeated application of the Lie brackets method. The appendices provide the mathematical details and the fundamentals of geometric control theory.

\section{Model}
\label{sec: Model}
\begin{figure}[t]
    \centering
\includegraphics[scale=0.7]{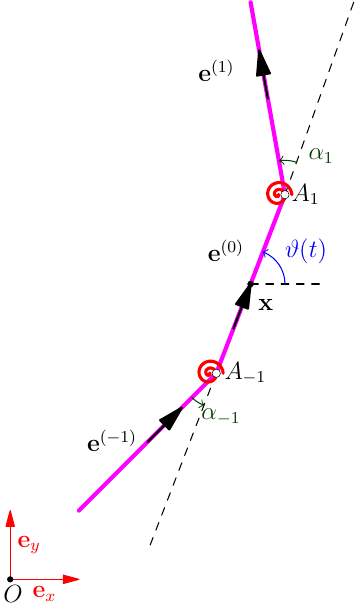}
    \caption{Schematic representation of an elastic $3$-links microswimmer.}
    \label{fig:model}
\end{figure}

This section derives the governing equations for a Purcell-type swimmer equipped with torsional springs at the joints (Figure \ref{fig:model}). The aim is to study the motion of the swimmer, controlled by external inputs dictating the desired joint angles. Unlike previous models, this approach does not directly control the shape angles but instead the target shape the swimmer should adopt over time.
We introduce a Cartesian coordinate system centred at point $O$. The unit vectors along the $x$- and $y$-axes are denoted as $\ev_x$ and $\ev_y$, respectively. Let $\xv$ represent the position vector of the midpoint of the microswimmer's central link. Its Cartesian components are given by $x(t)$ and $y(t)$, such that 
\be
\xv(t)=x(t) \ev_x + y(t) \ev_y. 
\ee
Finally, let $\vartheta$ represent the time-dependent angle between the central link and the positive $x$-axis.
Here, the subscript $j={-1,0,+1}$ denotes the microswimmer links. More precisely, $j = 0$ represents the central link, while $j = -1$ and $j = +1$ represent the lateral links. The orientation of each link is described by the planar unit vectors, 
\begin{equation}
    \label{links}
    \ej(t) :=  \cos \phij(t) \ev_x + \sin \phij(t)  \ev_y, 
\end{equation}
with 
\[
\phij(t)  := \vartheta(t) + j \alphaj(t), 
\]
with $\alpha_{\pm 1}$ denoting the alignment angles between adjacent links, as shown in Figure \ref{fig:model}. Thus, the triple ${x(t),y(t),\vartheta(t)}$ describes the position and orientation of the microswimmer in the plane, while the pair ${\alpha_{-1}(t), \alpha_{+1}(t)}$ defines its shape.

The swimmer's motion is driven by external forces from the fluid's reaction to its shape changes. According to resistive force theory \cite{10.1242/jeb.32.4.802}, the force and torque exerted by the fluid on each point of a slender body depend solely on the local velocity at that point, neglecting interactions with neighbouring points. Consequently, we can express the force density on the $j$-th link as
\be
\fj(s,t)=-\xi \; {v}_{j \parallel} (s,t) \ej(t) -\eta \; {v}_{j \perp}(s,t) \nj(t),
\ee
where  $\nj (t) := \ev_z \times \ej(t)$ and
 \be
 {v}_{j \parallel}(s,t) := \vj(s,t)\cdot\ej(t),  \quad  {v}_{j \perp}(s,t) := \vj(s,t)\cdot\nj(t),
 \ee
  represent the longitudinal and transverse components of the velocity, respectively, while $\xi$ and $\eta$ denote the positive drag coefficients along the directions $\ej$ and $\nj$, respectively.

The position of a point on the $j-$th link of the microswimmer can be formulated as
\be
 \label{eq: curvilinear ascissa}
 \xj(s,t) = \xv(t) + j \frac{L}{2} \ev^{(0)} (t) + \left(s + j \frac{L}{2}\right)\ej(t),
\nonumber
\ee
whence
\be
 \label{eq:velocities}
 \vj(s,t) = \dot \xv_j (t) + j \frac{L}{2} \dot \vartheta(t)  \nv_0 (t) + \left(s + j \frac{L}{2}\right)\phijd(t) \nj(t). 
\nonumber
\ee
that can be used to compute the total hydrodynamic force and torque according to:
\be
    \label{eq: hydrodynamic force}
    \hf(t)=\sum_{j=-1}^{1}\int_{-\frac{L}{2}}^\frac{L}{2} \fj(s,t) \de s, 
 \ee
 and
 \be   
    \Mv(t) =  \sum_{j=-1}^{1}\int_{-\frac{L}{2}}^\frac{L}{2}  \left(\xj(s,t)-\xv(t)\right) \times \fj(s,t) \de s.
\ee
In the low Reynolds number regime, inertial forces are negligible. Consequently, the two equations governing the swimmer's motion, corresponding to force and torque balance, are
\be
\Fv(t)= {\bf 0}, \qquad \Mv(t) = {\bf 0}.
\ee
Note that these two vector equations, which give three scalar equations, are necessary but not sufficient to fully describe the swimmer's motion, which has five degrees of freedom. Moreover, they do not account for the effects of the torsional springs, which are internal torques exchanged between links of the swimmer and do not contribute to the balance of external forces and torques.

Two additional scalar equations, which do not introduce any new unknowns, are derived from the torque balance on each lateral link at its respective joint. For each of the two links, the total torque includes the hydrodynamic torque acting on the lateral link and the torque exerted by the torsional spring at the joint. Thus, denoting by $\kappa$ the stiffness of the torsional springs, we get
\begin{eqnarray}
\Mv_j(t) =  \int_{-\frac{L}{2}}^\frac{L}{2} \left(s + j \frac{L}{2}\right)\ej(t) \times \fj (s,t) \de s 
- \kappa \left(\aj (t) - \bar\aj\right)\ev_z , \quad j=\pm 1
\label{eq:Mj}
\end{eqnarray}
where the angles $\bar\aj$ denote the {\it spontaneous alignment angles}, i.e., the angles the swimmer would assume in the absence of external forces. Thus, the presence of torsional springs results in a swimmer shape where $\alphaj = \bar\aj$. A key novelty of this study is considering the angles $\bar\aj$ as control gaits.
\subsection{Dimensionless equations of motion}
To render the equations of motion dimensionless, we define the characteristic time,
\begin{equation*}
\tau_c = \frac{\xi L^3}{\kappa},
\end{equation*}
obtained through dimensional analysis by comparing the torque exerted by the coil to the torque exerted by the fluid on a swimmer link \cite{passov2012dynamics}. Subsequently, all physical quantities are scaled by their respective characteristic scales:
\begin{equation}
x = x^* L, \qquad y = y^* L, \qquad t = t^* \tau_c
\end{equation}
with forces $\mathbf{F}$ and torques $\mathbf{M}$ scaling as
\begin{equation}
\mathbf{F} = \mathbf{F}^* \left(\frac{\rho L^2}{\tau_c}\right), \qquad \mathbf{M} = \mathbf{M}^* \left(\frac{\rho L^3}{\tau_c}\right). 
\end{equation}
 For the sake of simplifying the formalism, we will henceforth omit the asterisk symbol from all quantities. It should be understood that these variables represent scaled dimensionless quantities.
 
 Summarising, the equations of motion in dimensionless form are
\begin{equation}
\label{eq: system of motion}
\begin{cases}
    \ev_x \cdot  \Fv(t)=0,\\ 
    \ev_y \cdot  \Fv(t)=0,\\ 
    \ev_z \cdot \Mv(t) =0, \\
    \ev_z \cdot \Mv_{-1}(t) =0, \\
    \ev_z \cdot \Mv_{+1}(t) =0.
\end{cases}
\end{equation}
for the five unknowns $x(t)$, $y(t)$, $\vartheta(t)$, $\am(t)$ e $\ap(t)$. 

Using  equations \eqref{eq: hydrodynamic force} and \eqref{eq:Mj} and introducing  the state vector 
\[
\yv(t) :=[x(t), y(t), \vartheta(t), \am(t),\ap(t)]^T,
\]
 equation \eqref{eq: system of motion} can be recast in the form:
\be
{\mathcal{R}}(t) \dot \yv(t) = \bv(t),
\label{eq:stato}
\ee
where $\mathbf{\mathcal{R}}(t)$ is $5 \times 5$  symmetric matrix  \cite{alouges2015can}, whose entries are detailed in Appendix \ref{R}, and
\be
\bv(t) = [0,0,0,\am(t) - u_1(t), \ap(t)-u_2(t)].
\label{eq:b}
\ee
Note that in equation \eqref{eq:b}, $u_1$ and $u_2$ denote the angles $\bam$ and $\bap$, respectively, which will serve as control inputs.

Following \cite{alouges2017purcell}, since the first two equations of system \eqref{eq: system of motion} do not depend on $x(t)$ and $y(t)$, the system is coupled in one direction only, with the orientation variables  $\vartheta(t)$, , $\alpha_{-1}(t)$ e $\alpha_{+1}(t)$  
influencing the translational variables $x(t)$ and $y(t)$, but not vice versa.

In fact, the matrix $\mathcal{R}(t)$ can be decomposed into blocks as follows
\begin{equation}
    \mathcal{R}(t) = \begin{bmatrix}
        \mathbf{A} & \mathbf{B} \\
        \mathbf{B}^{T} & \mathbf{C}
    \end{bmatrix},
\end{equation}
where $\mathbf{A}$ $\mathbf{B}$ and $\mathbf{C}$ are respectively $2 \times 2$, $2\times 3$ and $3\times 3$ matrices,  all of which depend only on the variables $\vartheta$, $\alpha_{-1}$ $\alpha_{+1}$.
Thus, the equations of motion can be be decoupled into two subsystems:
\begin{subequations}
\begin{equation}
\label{x_sy}
\begin{bmatrix}
\dot x\\
\dot y\\
\end{bmatrix}
 = - \mathbf{A}^{-1} \mathbf{B}
\begin{bmatrix}
\dot{\vartheta} \\
\dot{\alpha}_{-1} \\
\dot{\alpha}_{+1}
\end{bmatrix}
\end{equation}
\begin{equation}
\label{eq: orientation and shape subsystem}
\left( -\mathbf{B}^T \mathbf{A}^{-1} \mathbf{B} + \mathbf{C} \right)
\begin{bmatrix}
\dot{\vartheta} \\
\dot{\alpha}_{-1} \\
\dot{\alpha}_{+1}
\end{bmatrix}
= \begin{bmatrix}
0 \\
\alpha_{-1} -u_1\\
\alpha_{+1}-u_2
\end{bmatrix}
\end{equation}
\end{subequations}
This allows us to integrate the equations of motion in two steps. First, we solve system \eqref{eq: orientation and shape subsystem}, and then use its solutions to solve system \eqref{x_sy}.

\section{Small-amplitude oscillating controls}
\label{sec:2}
Inspired by previous research \cite{passov2012dynamics, alouges2017purcell}, we investigate the motion of a 3-link elastic microswimmer, assuming small oscillating equilibrium angles as control inputs. This allows us to assume consistently small alignment angles and, consequently, linearize the equations of motion for these angles. However, to accurately determine the displacement, we need to consider higher-order approximations of the equations of motion. Despite this, the resulting displacement predictions align well with numerical simulations obtained from the full nonlinear equations of motion. Furthermore, they are consistent with the findings of \cite{zigelman2024dynamics}.

\subsection{Linear analysis}
\label{sec: Anlytic analysis}
We consider the equilibrium point \\
$((\vartheta,\alpha_{-1},\alpha_{+1}),(u_1,u_2))=((\vartheta_0,0,0),\,(0,0))$ and assume  small-amplitude periodic control gaits 
\begin{equation}
\label{osc_contr}
u_1(t)=\epsilon\sin(\omega t), \qquad u_2(t)=\epsilon \sin(\omega t+\phi),
\end{equation}
with $\epsilon \ll1 $ the dimensionless amplitude, $\omega$ the dimensionless frequency, and $\phi$ a shift phase. 

Consistently, we assume the unknown variables can be expanded in a power series of $\epsilon$:
\begin{subequations}
\begin{equation}
    x(t) =  \sum_{k=1}^\infty \epsilon^k x^{(k)}(t),
\quad 
    y(t) =  \sum_{k=1}^\infty \epsilon^k y^{(k)}(t),
\end{equation}
\begin{equation}
    \vartheta(t) = \vartheta_0 + \sum_{k=1}^\infty \epsilon^k \vartheta^{(k)}(t),
\quad 
    \alpha_{\pm 1}(t) =  \sum_{k=1}^\infty \epsilon^k \alpha_{\pm1}^{(k)}(t).
\end{equation}
\label{eq:approx}
\end{subequations}
Replacing equations \eqref{eq:approx} into \eqref{eq: orientation and shape subsystem},
$O(\epsilon)$ we obtain
\begin{equation}
\label{eq: angles' system}
    \begin{bmatrix}
        \dot \vartheta^{(1)}\\
        \dot \alpha_{-1}^{(1)}\\
        \dot \alpha_{+1}^{(1)}
    \end{bmatrix} 
     = {\nu}
\begin{bmatrix}
 d & c & c \\
 c & a & b \\
 c & b & a \\
\end{bmatrix}
 \begin{bmatrix}
0 \\
\alpha^{(1)}_{-1} - \sin(\omega t)\\
\alpha^{(1)}_{+1} - \sin(\omega t+\phi)
\end{bmatrix},
\end{equation}
where $\nu=\xi/\eta$, $a=-96/5$, $b=-66/5$, $c=42/5$  and $d=-24/5$. 

Solving the system \eqref{eq: angles' system}, we get 
%\begin{widetext}
\begin{subequations}
\label{lin_sol}
\begin{eqnarray}
\label{eq:t1} 
    \vartheta^{(1)}(t)= c_3 + \frac{c(c_1 + c_2)}{a+b} \left({\rm e}^{(a+b)\nu t}-1\right) + \frac{2 c \nu}{\nu ^2 (a+b)^2+\omega ^2}\times\\
    \times{ \left(\nu  (a+b) \sin \left(\omega t
   +\frac{\phi }{2}\right)+\omega  \cos \left(\omega t +\frac{\phi
   }{2}\right)\right)} \cos \left(\frac{\phi }{2}\right),
   \nonumber
\end{eqnarray}
\begin{eqnarray}
\label{eq:am}
\alpha^{(1)}_{-1}(t)= {\rm e}^{a \nu t}\left(c_1\cosh(b \nu t)+c_2 \sinh(b \nu t)\right)+\\+ q_1 \cos(\omega t) + q_2 \cos(\omega t + \phi) + q_3 \sin(\omega t) + q_4 \sin(\omega t + \phi),  \nonumber
    \end{eqnarray}
\begin{eqnarray}
\label{eq:ap}
    \alpha^{(1)}_{+1}(t)= {\rm e}^{a \nu t}\left(c_1\sinh(b \nu t)+c_2 \cos(b \nu t)\right) +\\+ q_1 \cos(\omega t + \phi) + q_2 \cos(\omega t) + q_3 \sin(\omega t + \phi) + q_4 \sin(\omega t),  \nonumber
    \end{eqnarray}
\end{subequations}
%\end{widetext}
where $c_i$, $i=1,2,3$, are integration constants, and
\[
\begin{aligned}
q_1&= \frac{a \nu  \omega  \left(\nu ^2 (a^2 - b^2)+\omega ^2\right)}{\nu ^4
   \left(a^2-b^2\right)^2+2 \nu ^2 \omega ^2 \left(a^2+b^2\right)+\omega ^4},
\\
q_2&= \frac{b \nu  \omega  \left(\nu ^2 \left(b^2-a^2\right)+\omega ^2\right)}{\nu ^4 \left(a^2-b^2\right)^2+2 \nu ^2 \omega ^2 \left(a^2+b^2\right)+\omega
   ^4},
\\
q_3&=\frac{\nu ^4 \left(a^2-b^2\right)^2+\nu ^2 \omega ^2 \left(a^2+b^2\right)}{\nu ^4
   \left(a^2-b^2\right)^2+2 \nu ^2 \omega ^2 \left(a^2+b^2\right)+\omega ^4},
\\
q_4&=\frac{2 a b \nu ^2 \omega ^2}{\nu ^4 \left(a^2-b^2\right)^2+2 \nu ^2 \omega ^2
   \left(a^2+b^2\right)+\omega ^4}.
   \end{aligned}
\]
Expanding equation \eqref{x_sy}  to first order in  $\epsilon$ yields
\begin{equation}
\begin{bmatrix}
    \dot x^{(1)}\\
    \dot y^{(1)}
\end{bmatrix} = \frac{1}{6}
    \left(
\begin{array}{ccc}
 0 & -\sin \vartheta_0 & \sin  \vartheta_0 \\
 0 & \cos  \vartheta_0 & -\cos  \vartheta_0 \\
\end{array}
\right)
\begin{bmatrix}
    \dot \vartheta^{(1)}\\
    \dot \alpha_{-1}^{(1)}\\
    \dot \alpha_{+1}^{(1)}
\end{bmatrix}, 
\end{equation}
which easily lead to
\begin{equation}
x^{(1)}(t) =   \frac{\ap - \am}{6} \sin \vartheta_0,
\qquad 
y^{(1)}(t) = \frac{\am - \ap}{6} \cos \vartheta_0,
\label{eq:sp1}
\end{equation}
These equations indicate that, to leading order, the trajectory of the centre of the central link is orthogonal to the link's direction. Note that, in equations \eqref{eq:am} and \eqref{eq:ap}, $a$ and $b$ are both negative with $a < b$. This implies that the part of the solution involving arbitrary constants is damped. Consequently, for sufficiently long times, only the periodic solution persists. As a result, the trajectory described by equations \eqref{eq:sp1} also becomes periodic for sufficiently long times, resulting in no net advancement of the swimmer.

\subsection{Weakly nonlinear analysis}
To determine whether the swimmer experiences net displacement under the influence of small-amplitude oscillating controls, it is essential to analyse higher-order terms in the perturbation expansion.

We first consider equation \eqref{eq: orientation and shape subsystem} that, at $O(\epsilon^2)$, leads to the linear homogeneous system
\begin{equation}
%\label{eq: angles' system}
    \begin{bmatrix}
\dot{\vartheta}^{(2)} \\
\dot{\alpha}^{(2)}_{-1} \\
\dot{\alpha}^{(2)}_{+1}
\end{bmatrix} = \nu
\begin{bmatrix}
0 & c & c \\
 0 & a & b \\
 0 & b & a \\
\end{bmatrix}
\begin{bmatrix}
{\vartheta}^{(2)} \\
{\alpha}^{(2)}_{-1} \\
{\alpha}^{(2)}_{+1}
\end{bmatrix}
\end{equation}
whose admit only the damped solution. However, when it is solved  with null initial conditions, it gives the trivial solution
\begin{equation}
\vartheta^{(2)}(t)=0, \qquad  {\alpha}^{(2)}_{-1}(t)=0, \qquad {\alpha}^{(2)}_{+1}(t)=0.
\end{equation}
This implies that, to order $O(\epsilon^2)$, the right-hand side of equation \eqref{x_sy} depends only on the variables $\vartheta^{(1)}, \alpha^{(1)}_{-1}, \alpha^{(1)}_{+1}$:
\begin{equation}
\label{eq:xyO2}
\begin{bmatrix}
\dot x^{(2)}\\
\dot y^{(2)}
\end{bmatrix}
= 
(a_{11}\dot\vartheta^{(1)} + a_{12} \dot\alpha^{(1)}_{-1} +  a_{13} \dot\alpha^{(1)}_{+1})
\begin{bmatrix}
\cos \vartheta_0\\
\sin \vartheta_0
\end{bmatrix},
\end{equation}
where 
\[
a_{11} :=  \frac{(\nu - 2)(\am^{(1)} - \ap^{(1)})}{6 \nu},
\quad
a_{12} := -\frac{(2  + \nu)\am^{(1)} + (\nu - 1)\ap^{(1)} + 3 \nu \vartheta^{(1)} }{18 \nu},
\]
\[
a_{13} := \frac{ (\nu - 1)\am^{(1)} + (2  + \nu)\ap^{(1)} + 3 \nu \vartheta^{(1)} }{18 \nu}.
\]
Unfortunately, due to the nonlinear nature of system \eqref{eq:xyO2}, an exact analytical solution cannot be obtained. However, following the approach described in \cite{alouges2017purcell}, we can focus on the long-term behaviour and compute the net displacement of the microswimmer over one period. Thus, by substituting the periodic components of the solutions (\ref{eq:t1}-\ref{eq:ap}) into equation \eqref{eq:xyO2} and integrating over a period  $T=2 \pi/\omega $, we obtain:
\begin{equation}
\label{spostamento}
\begin{bmatrix}
 \Delta x^{(2)}\\
 \Delta y^{(2)}
\end{bmatrix}
= \frac{60 \pi  (\nu -1) \nu  \left(972 \nu ^2+5 \omega ^2\right) \sin \phi }{\left(36 \nu
   ^2+\omega ^2\right) \left(26244 \nu ^2+25 \omega ^2\right)}
\begin{bmatrix}
\cos \vartheta_0\\
\sin \vartheta_0
\end{bmatrix}.
\nonumber
\end{equation}
Therefore,  the magnitude of the displacement  after one period  of the periodic solution is: 
 \begin{equation}
 \label{d}
 \Delta^{(2)}= \epsilon^2 \frac{60 \pi   \nu  \left(972 \nu ^2+5 \omega ^2\right) (1- \nu) |\sin \phi| }{\left(36 \nu
   ^2+\omega ^2\right) \left(26244 \nu ^2+25 \omega ^2\right)},
 \end{equation}
along the $\vartheta_0$ direction.  The maximum displacement is achieved for a phase shift of $\phi = \pm {\pi}/{2}$. Moreover, opposite signs of $\phi$ correspond to opposite directions of the microswimmer's motion. 

\subsection{Asymptotic leading order solution}
Below we summarize the long-term solutions obtained in the previous sections.
Since the maximum displacement over a period occurs for $\phi=\pm \pi/2$, we fix $\phi = \pi/2$.  The asymptotic motion of the centre of the central link is obtained by combining the periodic parts of the solutions \eqref{eq:sp1} and \eqref{eq:xyO2}, obtaining:
\be
\begin{bmatrix}
  x^\infty(t)\\
  y^\infty(t)
\end{bmatrix}
=  \epsilon
\sigma^{(1)}(t)
\begin{bmatrix}
  -\sin \vartheta_0 \\
  \cos \vartheta_0
\end{bmatrix}
+ \epsilon^2
\sigma^{(2)}(t)
\begin{bmatrix}
  \cos \vartheta_0 \\
  \sin \vartheta_0
\end{bmatrix},
\ee
where
%\begin{widetext}
\be
\sigma^{(1)}(t)  = \frac{\nu ((\omega -6 \nu ) \sin (\omega t)+(6 \nu +\omega ) \cos
   ( \omega t))}{36 \nu ^2+\omega ^2},
\ee
\be
\begin{aligned}
\sigma^{(2)}(t) =&\frac{1}{\left(36 \nu ^2+\omega ^2\right)(26244 \nu ^2+25 \omega^2)}\bigg(6 \nu  (5 (\nu -1) t \omega  (972 \nu ^2+5 \omega ^2))+\\
   & (9 \nu +4)
   \left(\left(5 \omega ^2-972 \nu ^2\right) \sin ^2(\omega t)+96 \nu  \omega  \sin (2 
   \omega t)\right)\bigg).
\end{aligned}
\ee
%\end{widetext}
This analysis reveals that the motion consists of two components: a primary periodic component orthogonal to the central link and a secondary component parallel to the link. Only the latter contributes to the net advancement of the swimmer over a period. As a result, the microswimmer undergoes large periodic transverse motions in order to achieve a small forward displacement. 

Furthermore,  throughout this phase, the direction of the central link fluctuates around its initial orientation:
\begin{equation}
\vartheta^\infty(t) = \vartheta_0- \epsilon \frac{42 \nu  ((162 \nu +5 \omega ) \sin (\omega t )+(162 \nu -5 \omega ) \cos (\omega t
))}{26244 \nu ^2+25 \omega ^2}.
\end{equation}
while the lateral links oscillate around the the strait configuration
%\begin{widetext}
\be
\begin{aligned}
\am^\infty(t) &=\frac{\epsilon}{D} \bigg(6 \nu ((157464 \nu ^3-10692 \nu ^2 \omega +2262 \nu  \omega ^2+55 \omega
   ^3) \sin (\omega t )-\\
   &16 \omega  \left(972 \nu ^2-132 \nu  \omega +5 \omega ^2\right)
   \cos (\omega t))\bigg),
\end{aligned}
\ee
\be
\begin{aligned}
\ap^\infty(t) &=\frac{\epsilon}{D}\bigg(6 \nu (16 \omega  \left(972 \nu ^2+132 \nu  \omega +5 \omega ^2\right) \sin (
   \omega t)+\\
   &\left(157464 \nu ^3+10692 \nu ^2 \omega +2262 \nu  \omega ^2-55 \omega ^3\right)
   \cos (\omega t))\bigg),
\end{aligned}
\ee
where,
$$
D:=944784 \nu ^4+27144 \nu ^2 \omega ^2+25 \omega ^4.
$$
%\end{widetext}

In conclusion, in light of our analysis, we can assert that the solutions for the shape angles rapidly converge to a closed loop, representing a periodic solution. Consequently, the orientation of the swimmer also converges to a harmonic solution. The correction to this behaviour is at least of order $\epsilon^3$. This indicates that, on average, the swimmer moves along a straight trajectory, dictated by the initial orientation of the central link, without any net rotation.

Regarding the motion of the microswimmer, the variables $x(t)$ and $y(t)$ also exhibit similar behaviour: the solutions at $O(\epsilon)$ converge to harmonic solutions. However, the corrections at $O(\epsilon^2)$ indicate that the microswimmer experiences a net displacement over a cycle, even though the other variables follow closed loops.

Swimmer trajectories are shown in Figure \ref{fig:xytrajectory}, $\vartheta(t)$ and the shape angles trajectories in Figure \ref{fig:thetatrajectory}. The asymptotic harmonic solution is compared with the results of numerical simulations of the nonlinear model for various values of the amplitude $\epsilon$. The asymptotic approximation proves to be highly satisfactory, even for relatively large values of the parameter $\epsilon$.

\subsection{Comparison with numerical simulations}
The quasi-linear approximation captures only the leading-order motion of the swimmer with respect to the input amplitude $\epsilon$. To assess the accuracy of this approximation, we will compare certain asymptotic quantities with those obtained from numerical solutions of the nonlinear equations. The parameters used for all numerical integrations performed in this section are $\nu=\frac{1}{2}$ and $\vartheta_0=0$ with $\phi=\frac{\pi}{2}$.

A key quantity to compare, which helps us determine the microswimmer's performance as a function of frequency, is the average speed in the harmonic regime, given by
\begin{equation}
\label{eq:v_media}
\bar{v}=  \frac{\omega \Delta ^{(2)}}{2 \pi}.
\end{equation}
This is obtained by integrating equation \eqref{d} over a single period and dividing by the period length.

The average speed attains its maximum value at:
\be
 \omega_{\rm opt} = \pm {18} \sqrt{\frac{3}{5}}\frac{\xi}{\eta},
\ee
where it assumes the value: 
\be
\bar {v}_{\text{max}} = \frac{15 \sqrt{15}}{512} \left(1 - \frac{\xi}{\eta} \right) \epsilon^2.
\ee

\begin{figure}
    \centering
    \includegraphics[scale=0.7]{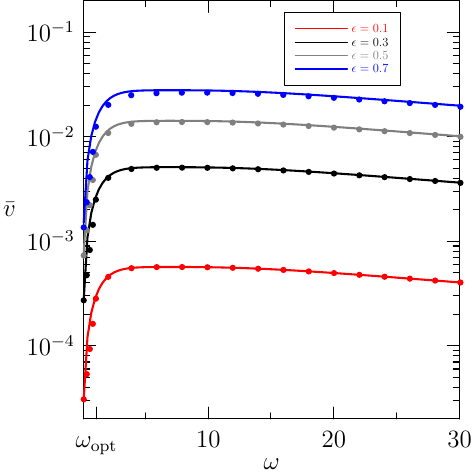}
    \includegraphics[scale=0.7]{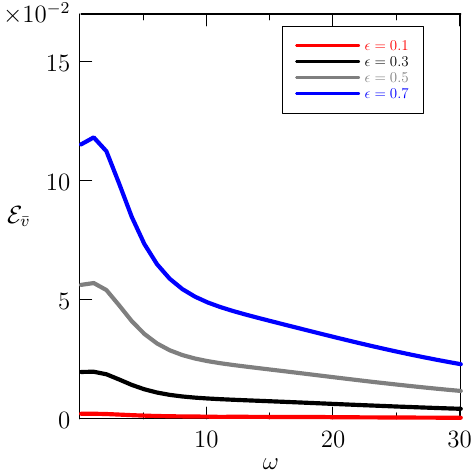}
    \caption{(Left) Plots of $\bar {v}$ versus frequency  $\omega$, are shown for $\varepsilon=0.1, 0.3, 0.5, 0.7$. Both theoretical results (solid lines) and numerical results (dotted lines) are included. (Right) 
    The relative error of the mean velocity,  $\mathcal{E}_{\bar{v}}$, quantifies the discrepancy between the theoretical and the numerical results.}
    \label{fig:v_media}
\end{figure}

Figure \ref{fig:v_media} (left) presents a comparison between the numerical solution of the full nonlinear problem and the analytical expression given by equation \eqref{eq:v_media}. The numerical value was calculated considering the long-time solution, i.e., in the harmonic regime, and averaging over 100 periods. The use of a large number of periods is particularly important for low frequencies, where the definition of average velocity can be unreliable due to significant fluctuations in displacement over a single period. Averaging over many periods mitigates this discrepancy. Figure \ref{fig:v_media} (right) illustrates the trend of the relative error $\mathcal{E}_{\bar{v}}$
\[ 
\mathcal{E}_{\bar{v}}=
\frac{|\bar{v}-\bar{v}^{\rm num}|}{|\bar{v}^{\rm num}|},
\] 
as a function of frequency $\omega$, for several values of the amplitude actuation $\epsilon$.

{In the following, we present a comparison between the results of numerical integration of the fully non-linear system and the theoretical predictions obtained from linear and weakly non-linear analyses. For these simulations, we set $\omega = \omega_{\text{opt}}$.}

\begin{figure}
    \centering
    \includegraphics[scale=0.7]{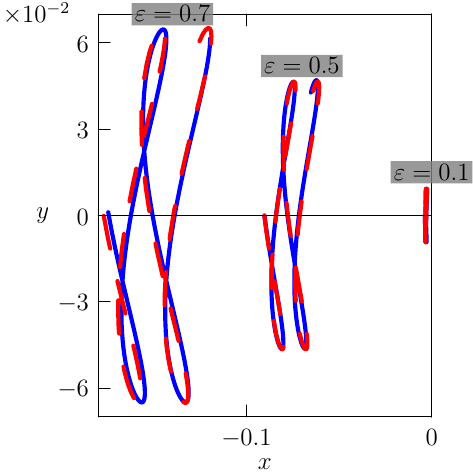}
    \caption{Trajectories in the $x-y$ plane for different values of $\epsilon$. Solid blue lines represent the numerical solutions of the fully nonlinear equations of motion, while dashed red lines represent solutions obtained from the weak nonlinear analysis.}
    \label{fig:xytrajectory}
\end{figure}

\begin{figure}
    \centering
    \includegraphics[scale=0.6]{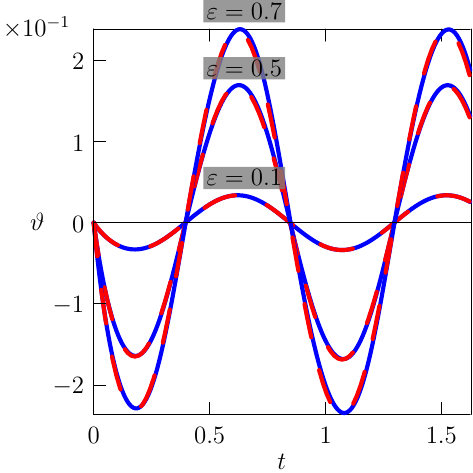}
    \includegraphics[scale=0.6]{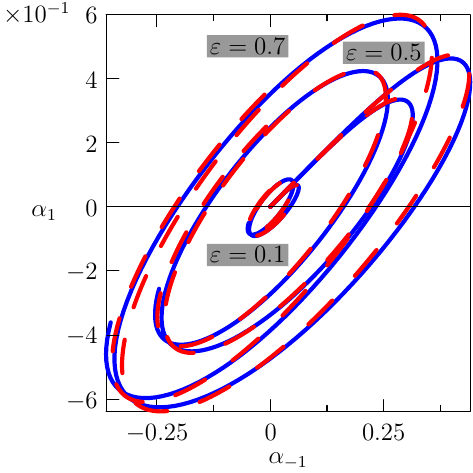}
    \caption{Orientation angle $\vartheta$ as a function of time (left) and parametric plot of the shape angles in the $\alpha_{-1}- \alpha_{+1}$ plane (right), are shown for several values of the amplitude control gaits $\epsilon$.  Solid blue lines represent the numerical solutions of the full nonlinear equations of motion, while dashed red lines represent solutions obtained from linear analysis}
    \label{fig:thetatrajectory}
\end{figure}

Figure \ref{fig:xytrajectory} illustrates the trajectory of the studied microswimmer in the $x-y$ plane. To explore the effect of varying control amplitudes, the trajectory is plotted for different values of the actuation amplitude $\epsilon$. For $\epsilon=0.1$, the plot shows the first two periods of integration. For $\epsilon=0.5$, both trajectories are plotted from $t=4\pi/\omega$, while for $\epsilon=0.7$, both trajectories are plotted starting at $t= 6\pi/\omega$. This choice was made to ensure that the different trajectories are clearly distinguishable within the same figure. The close alignment between these two sets of results demonstrates excellent agreement, thereby validating the theoretical approach.
Moreover, the numerical results exhibit excellent agreement with our asymptotic theoretical predictions for the variables  $\vartheta(t)$ and $\alpha_{\pm 1}(t)$.  As shown in Figure \ref{fig:thetatrajectory}, the dashed red lines, representing the theoretical predictions, closely follow the blue lines, which depict the curves obtained from numerical integration.
\section{Small-Time Local Controllability}
\label{sec:Controllability}
Linearisation around the equilibrium point provides valuable insights into the system's dynamics, revealing that the orientation and shape angles converge to steady periodic solutions with short periods over extended time intervals. Furthermore, the linear analysis indicates that the system exhibits translational behaviour in the direction of $\vartheta$. Building upon this understanding, we now aim to investigate the controllability of the system, specifically determining whether it can be steered to any desired point within a neighbourhood of the equilibrium.

Returning to equation of motion \eqref{eq:stato} and noting that ${\mathcal{R}}$ is invertible \cite{alouges2015can}, we can define:
\[
\qv_0(\mathbf{y}):=  \mathbf{\mathcal{R}}^{-1}(t) [0,0,0,\am(t),\ap(t)]^T,
 \]
 \[
 \qv_1(\mathbf{y}) := \mathbf{\mathcal{R}}^{-1}(t)[0,0,0,-1,0]^T,
\quad
  \qv_2(\mathbf{y}) := \mathbf{\mathcal{R}}^{-1}(t)[0,0,0,0,-1]^T,
 \]
and rewrite the equations of motion as a nonlinear affine control system with drift: 
\begin{equation}
    \label{eq:control affine}
    \dot{\mathbf{y}}= \qv_0(\mathbf{y}) + \sum_{k=1}^2 u_k \qv_k(\mathbf{y}),
\end{equation}
with control gaits $u_k(t)$.

The study of controllability for nonlinear affine control systems, as defined by equation \eqref{eq:control affine}, is challenging due to their inherent nonlinearity. Therefore, we will focus on small-time local controllability. To this end, we will require several concepts from geometric control theory, including the definition of small-time local controllability itself. For a more detailed discussion, we refer the reader to Coron's book \cite{coron2007control}. However, we will provide a concise summary of the relevant definitions and theorems in Appendix \ref{appendiceB}.

\subsection{STLC of the elastic microswimmer}
\label{Double Input}
To demonstrate the small-time local controllability of system \eqref{eq:control affine} with $u_1(t) \neq u_2(t)$ for $t\geq 0$, 
we consider the equilibrium point  $(\mathbf{y}_e,\mathbf{u}_e)$, where $\mathbf{y}_e=[x,y,\vartheta_0,0,0]$
and $\mathbf{u}_e=[0,0]$, corresponding to a straight microswimmer.  We will employ Sussmann's condition to verify controllability.
\begin{theorem}
    System \eqref{eq:control affine} is small time locally controllable around  equilibrium points ~$(\mathbf{y}_e,\mathbf{u}_e)=((x,y,\vartheta_0,0,0),(0,0))$.
\end{theorem}
\textbf{proof}
The proof relies on Sussmann's condition and Theorem \ref{01}  in Appendix \ref{appendiceB}.

First, we consider the Lie algebra generated by the vector fields, $ {\rm Lie}(\{\qv_0,\qv_1,\qv_2\})$. The Lie algebra rank condition requires that the Lie brackets of these vector fields span the entire tangent space at the equilibrium point. To verify this, we define the following vector fields:
\label{eq:spanvectors}
    \begin{equation}
    \label{g1}\gv_1(\mathbf{y}_e)=\qv_1(\mathbf{y}_e),
   \quad
         \gv_2(\mathbf{y}_e)=\qv_2(\mathbf{y}_e),
 \quad
         \gv_3(\mathbf{y}_e)=[\qv_1,\qv_2](\mathbf{y}_e), \nonumber
    \end{equation}
    \begin{equation}
         \gv_4(\mathbf{y}_e)=[\qv_1,[\qv_1,\qv_2]](\mathbf{y}_e),
   \quad
           \gv_5(\mathbf{y}_e)=[\qv_2,[\qv_1,\qv_2]](\mathbf{y}_e), \nonumber
    \end{equation}
By calculating the explicit expressions for these vector fields, we can built explicitly  the matrix  $\Lv = (\gv_1, \gv_2, \gv_3,\gv_4,\gv_5)$: 
%\begin{widetext}
 \begin{equation}
    \mathbf{L}(\mathbf{y}_e)=-\nu
\begin{bmatrix}
\sin{\vartheta_0} & -\cos{\vartheta_0} & c & a & b \\
- \sin{\vartheta_0} & \cos{\vartheta_0} & c & b & a \\
{12  (\nu-1 ) \cos{\vartheta_0}} & 12  (\nu-1 ) \sin{\vartheta_0} & 0 & 0 & 0 \\
 -\nu l_a \sin{\vartheta_0} & \nu l_a \cos{\vartheta_0} & %\frac{1188 \xi ^2 \left(\eta ^2+8 \xi ^2\right)}{25 \eta ^4} 
 -\nu l_b
 & -\nu l_c & \nu l_d \\
-\nu l_a \sin{\vartheta_0}
& \nu l_a \cos{\vartheta_0}
& -\nu l_b
%-\frac{1188 \xi ^2 \left(\eta ^2+8 \xi ^2\right)}{25 \eta ^4} 
& -\nu l_d & \nu l_c 
\end{bmatrix},
\end{equation}
%\end{widetext}
where
\begin{subequations}
\begin{equation}
     %l_a:=\frac{18 \xi ^2 \left(95 \eta ^2+112 \eta  \xi +360 \xi ^2\right) \sin{\vartheta}}{25 \eta ^4},
     l_a:=\frac{57}{7}c-\frac{21}{5} a\nu -54 d \nu^2,
\qquad
    l_b:= -\frac{18}{5}b+\frac{3}{2}ab\nu^2,
\end{equation}
\begin{equation} 
l_c:=%\frac{864 \xi ^2 (2 \eta -\xi ) (5 \eta +8 \xi )}{25 \eta ^4},
-9a+\frac{3}{2}ab \nu +3ad\nu^2,
\qquad
l_d:=
%\frac{108 \xi ^2 \left(91 \eta ^2+128 \eta  \xi +244 \xi ^2\right)}{25 \eta ^4}.
-\frac{39}{4}cd+3ad\nu+\frac{183}{4}d^2\nu^2.
\end{equation}
\end{subequations}
Since $ \det \mathbf{L}$ vanishes at the real values $\eta/\xi=1$ and $\eta/\xi=-45/19$, that are not physically relevant for slender swimmers since $\eta/\xi>1$, we conclude that $\Lv$ has full rank. Hence, the vectors $\gv_i$ span the entire space $\mathbb{R}^5$, satisfying the Lie algebra rank condition.
 
To demonstrate that the Sussmann condition holds for all $\Lambda \in [0, 1]$, consider a generic element $\mathbf{h} \in \text{Br}(\{\qv_0, \qv_1, \qv_2\})$. We must select $\mathbf{h}$ such that $\delta_0(\mathbf{h})$ is odd and $\delta_i(\mathbf{h})$ is even for $i = 1, 2$.

The minimum total degree, denoted by $d$, for which such brackets exist is $d = 3$. When $d = 1$, the Sussmann condition trivially holds for all $\Lambda$ by choosing $\gv = \qv_1$ or $\gv = \qv_2$.

For brackets of degree $d = 3$, where $\qv_0$ appears an odd number of times and other $\qv_k$ appear an even number of times, $\mathbf{h}$ must take the form:
\begin{equation}
\label{eq:h}
\mathbf{h} = [\qv_k, [\qv_k, \qv_0]],
\end{equation}
for $k = 1, 2$.

Since it can be explicitly verified that
$
[\qv_1, [\qv_1, \qv_0]](\mathbf{y}_e) = -[\qv_2, [\qv_2, \qv_0]](\mathbf{y}_e),
$
we obtain
\[
\sigma(\mathbf{h})(\mathbf{y}_e) = [\qv_1, [\qv_1, \qv_0]](\mathbf{y}_e) + [\qv_2, [\qv_2, \qv_0]](\mathbf{y}_e) = \mathbf{0}.
\]
This implies that $\sigma(\mathbf{h})(\mathbf{y}_e)$ lies in the span of $\gv_1(\mathbf{y}_e)$. Since:
\begin{equation}
\label{g1Sus}
1 = \Lambda \delta_0(\gv_1) + \sum_{k=1}^2 \delta_k(\gv_1) < \Lambda \delta_0(\mathbf{h}) + \sum_{k=1}^2 \delta_k(\mathbf{h}) = \Lambda + 2, \quad \forall \Lambda \geq 0,
\end{equation}
the Sussmann condition is satisfied for all $\Lambda \in [0, 1]$. Therefore, by Theorem \ref{01}, the system is small-time locally controllable.
\begin{rem}
For iterated Lie brackets of degree greater than 3, verifying the Sussmann condition is unnecessary. This is because the Lie algebra rank condition, satisfied by vector fields of order up to 3, guarantees that higher-order brackets can be expressed as linear combinations of lower-order brackets. Consequently, $\sigma(\mathbf{h})$ also lies within the span of these lower-order brackets.
\end{rem}
%
% \begin{rem}
% Note that the small-time local controllability of the full system \eqref{eq:control affine} implies the controllability of the orientation and shape subsystem \eqref{eq: orientation and shape subsystem}. This is crucial for trajectory tracking problems.
% \end{rem}
\subsection{Displacement approximation with piecewise constant controls \label{sec:bracket_approx}}
Having established small-time local controllability, we aim to approximate the system's displacement using piecewise constant controls. Such controls are particularly well-suited for approximating system behaviour, especially for short time intervals, within the framework of geometric control theory

Consider a small time interval $\tau > 0$ and a constant $\gamma > 0$. We define the following piecewise constant control inputs:
\begin{equation}
\label{eq:u}
u_1(t) = 
\begin{cases}
    0, & 0 \leq t \leq \tau, \\
    \gamma, & \tau < t \leq 2\tau, \\
    0, & 2\tau < t \leq 3\tau, \\
    -\gamma, & 3\tau < t \leq 4\tau.
\end{cases}
\qquad
u_2(t) = 
\begin{cases}
    \gamma, & 0 \leq t \leq \tau, \\
    0, & \tau < t \leq 2\tau, \\
    -\gamma, & 2\tau < t \leq 3\tau, \\
    0, & 3\tau < t \leq 4\tau.
\end{cases}
\end{equation}
%\begin{widetext}
    \begin{figure}[t]
    \centering
    \includegraphics[scale=0.5]{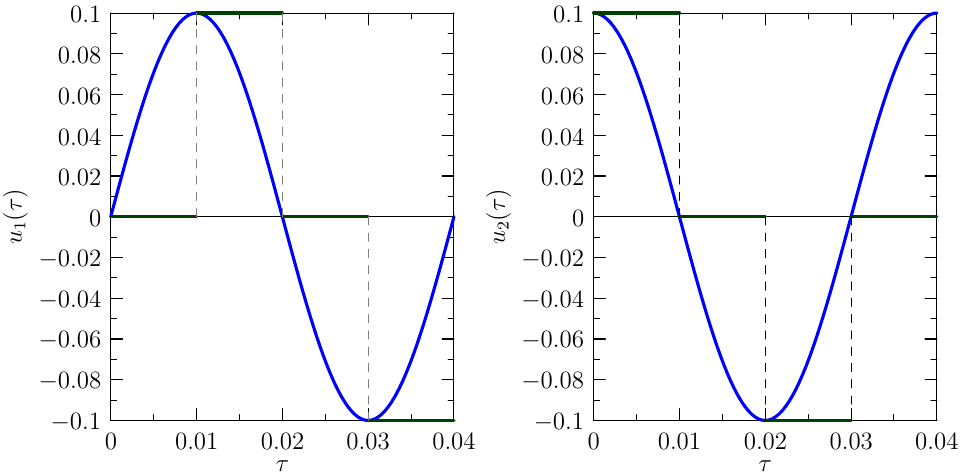}
    \caption{Continuous inputs with $\epsilon=0.1$ and their approximation with piecewise constant functions.}\label{fig:ApprossimazioneContinuiECostanti}
\end{figure}
 These control gaits, depicted in Figure \ref{fig:ApprossimazioneContinuiECostanti},  describe a periodic cycle of the swimmer's strokes, comprising four distinct phases. In each phase, the swimmer propels itself by activating only one of the two side links. A desired spontaneous configuration is assigned to each phase, which activates the torsional springs to push the link towards this configuration.
 
The variation of the state vector after one cycle is given by:
\begin{eqnarray}
\label{displacement}
\mathbf{y}(4\tau) = \mathbf{y}_0 + 4\tau \qv_0|_{\mathbf{y}_0} + \tau^2 \left(6(\nabla \qv_0 \cdot \qv_0) + 2\gamma [\qv_1, \qv_0]\right.  \nonumber \\ \left.
+ 2\gamma [\qv_2, \qv_0] - \gamma^2 [\qv_1, \qv_2]\right)_{\mathbf{y}_0} + o(\tau^2).%\nonumber
\end{eqnarray}
Considering the initial condition $\mathbf{y}_0 = \mathbf{y}_e$ and recalling that $\qv_0(\mathbf{y}_e) = \mathbf{0}$, we obtain:
\begin{eqnarray}
\label{eq: displacement in ye}
\mathbf{y}(4\tau) = 12 \gamma \nu \tau^2\Big({\gamma  (1 - \nu) \cos{\vartheta}},{ \gamma  (1 - \nu) \sin{\vartheta}},
  \nonumber \\  \frac{2268  \nu}{25}, -\frac{4374  \nu}{25}, -\frac{4374  \nu}{25}\Big) + o(\tau^2). 
  \label{eq:y}
\end{eqnarray}
Consequently, the relative errors ${\mathcal{E}}_x$, ${\mathcal{E}}_y$, $\mathcal{E}_{\vartheta}$, $\mathcal{E}_{\alpha_{-1}}$ and $\mathcal{E}_{\alpha_{+1}}$  between our theoretical prediction and the numerical integration are expected to be of the order  $o(\tau^2)/\tau^2\approx o(\tau)$,  since the difference between the theoretical and numerical displacements is of order $o(\tau^2)$ while the numerical displacement itself is expected to be of order $\tau^2$.

{The approximated displacement formula employing piecewise constant controls proves effective, as demonstrated by numerical simulations. Comparisons between the displacement obtained through numerical integration, using parameters $\xi=1, \, \eta=2, \, \gamma=0.1, \, \vartheta_0=0$,  and the theoretical prediction confirm the accuracy of this approach, as shown in Figure \ref{fig:x-error}}
%Figura 7
{\centering
\begin{figure}
    \includegraphics[scale=0.45]{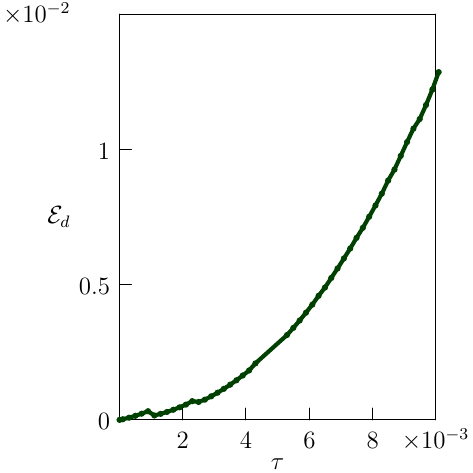}
    \includegraphics[scale=0.45]{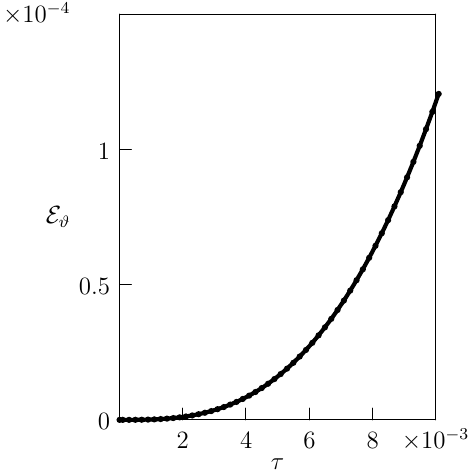}
\includegraphics[scale=0.45]{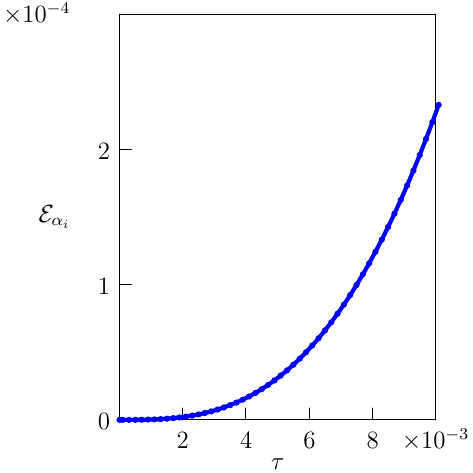}
    \caption{Relative error of the $x$-component defined as ${\mathcal{E}}_x:=\displaystyle\frac{\lvert x(4\tau)-x^{num}(4\tau)\rvert}{\lvert x^{num}(4\tau) \rvert}$, ${\mathcal{E}}_{\vartheta}:=\displaystyle\frac{\lvert \vartheta(4\tau)-\vartheta^{num}(4\tau)\rvert}{\lvert \vartheta^{num}(4\tau) \rvert}$, ${\mathcal{E}}_{\alpha_i}:=\displaystyle\frac{\lvert \alpha_i(4\tau)-\alpha_i^{num}(4\tau)\rvert}{\lvert \alpha_i^{num}(4\tau) \rvert}$, respectively}
    \label{fig:x-error}
\end{figure}}

\subsection{Comparison between continuous oscillating controls and piecewise constant
ones}
{
In this section, we highlight the similarities between the solution obtained by linearising the system and applying small-amplitude oscillating controls, and the solution obtained by iterating piecewise constant controls. Importantly, for small values of  $\gamma$, the controls in \eqref{eq:u} can be viewed as piecewise constant approximations of the continuous controls \eqref{osc_contr} (see Figure \ref{fig:ApprossimazioneContinuiECostanti}). Indeed, for small periods, the orientation and shape trajectories converge to periodic solutions, not only with continuous controls but also with their piecewise constant approximations. Moreover, in both cases, the trajectories in the plane asymptotically result in a net translation.}

{For the numerical simulations, we choose $\omega=\pi/(2\tau)$, with $\tau=0.01$. We approximate the oscillating controls \eqref{osc_contr} by the piecewise constant controls in  \eqref{eq:u} with $\gamma=\epsilon=0.1$, (see Figure \ref{fig:ApprossimazioneContinuiECostanti})}.

\begin{figure}
\centering
    \includegraphics[scale=0.6]{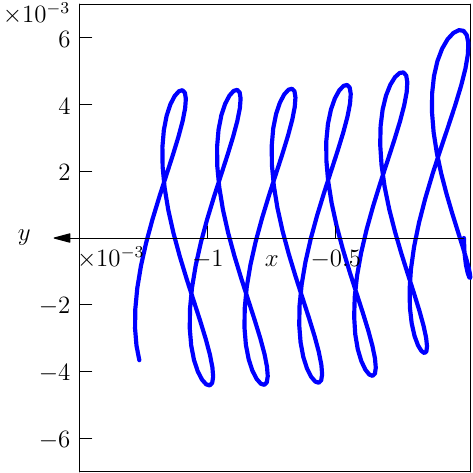}
    \includegraphics[scale=0.6]{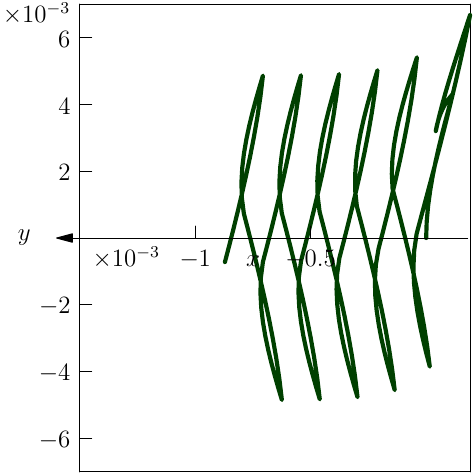}
    \caption{Numerical $x-y$ trajectory with continuous (left) and piecewise constant (right) inputs.}
    \label{fig:trajxy}
\end{figure}

\begin{figure}
    \centering
    \includegraphics[scale=0.6]{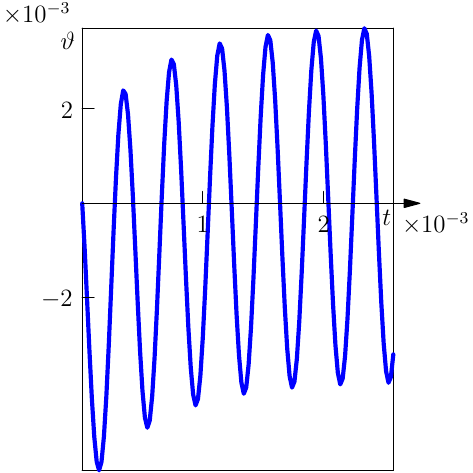}
    \includegraphics[scale=0.6]{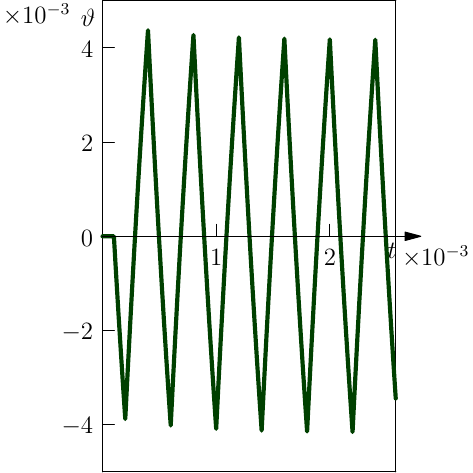}
    \caption{$\vartheta(t)$ with continuous (left) and piecewise constant (right) inputs.}
    \label{fig:tratheta}
\end{figure}

\begin{figure}
\centering
    \includegraphics[scale=0.6]{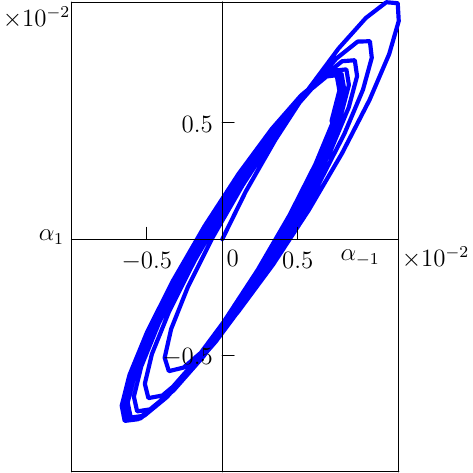}
    \label{fig:trajAlpha}
    \includegraphics[scale=0.6]{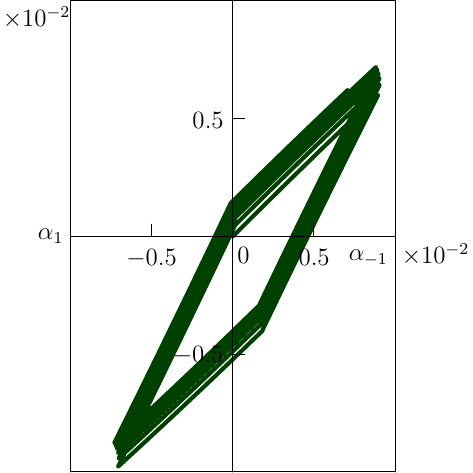}
    \caption{$\alpha_{-1}-\alpha_{+1}$ trajectory  with continuous (left) and piecewise constant (right) inputs.}
    \label{fig:trajAlpha1}
\end{figure}

{Upon integrating the equations of motion \eqref{eq:control affine} using either the oscillating continuous controls or the piecewise constant ones, extended by periodicity in the time interval   $[0,100 \, \tau]$,  with the initial condition $\mathbf{y}_e=(0,0,0,0,0)$, similar behaviour is observed:  Figure \ref{fig:trajxy} compares the $x-y$ trajectory, Figure  \ref{fig:tratheta} compares the variation of the orientation angle $\vartheta$, Figure \ref{fig:trajAlpha1} compares the shape angles. Indeed, these graphs demonstrate that for both input types, both the orientation angle and the shape angles converge towards a periodic solution while the system progresses along the orientation of the central link.}
\subsection{Iterated theoretical displacement}
In the following, we compare the evolution of the system obtained using two different methods: firstly, through numerical solution considering stepwise input gaits as defined in Equation \eqref{eq:u}; and secondly, by cumulating the quantities in a swimming cycle, obtained through the Lie bracket formalism (Section \ref{sec:bracket_approx}).

\begin{figure}
     \centering \includegraphics[scale=0.7]{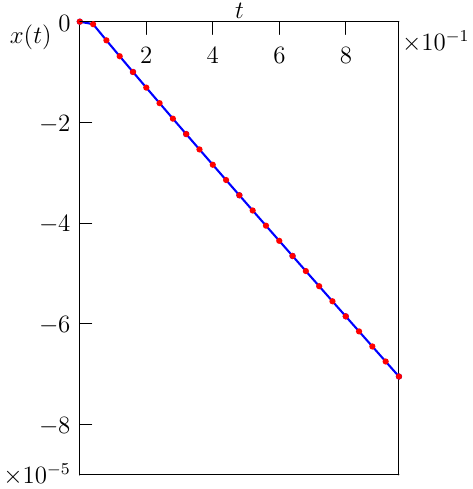}
     \includegraphics[scale=0.7]{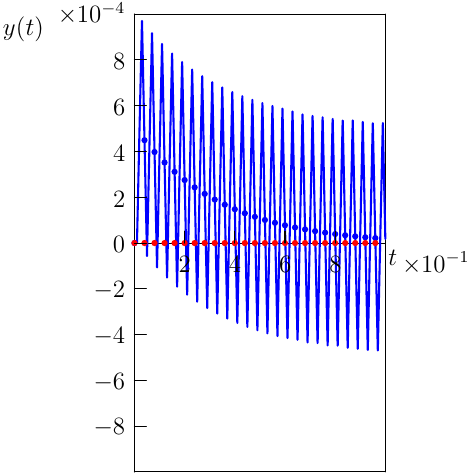}
    \caption{Comparison of the theoretical and numerical iterated displacements for the x-component (left) and y-component (right).}
    \label{fig:IteratedXDisplacement}
    \label{fig:IteratedYDisplacement}
\end{figure}

\begin{figure}
%     \begin{minipage}{0.45\textwidth}
\centering
\includegraphics[scale=0.45]{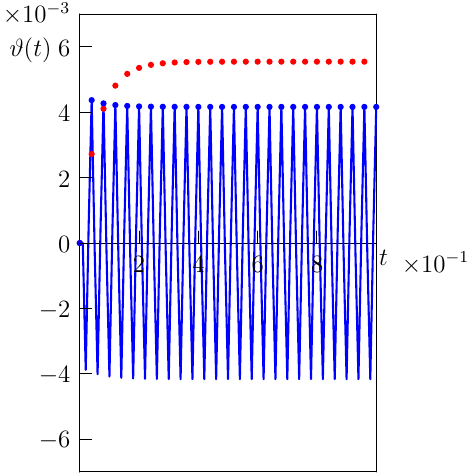}
\includegraphics[scale=0.45]{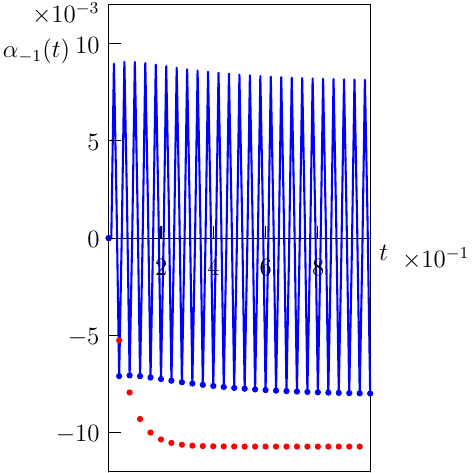}
\includegraphics[scale=0.45]{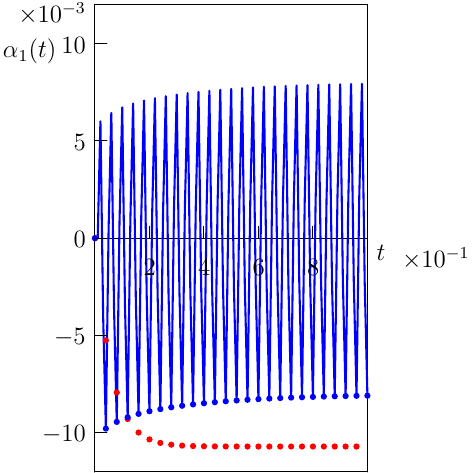}
\caption{Comparison between the theoretical and numerical iterated solutions for $\vartheta$(left), $\alpha_{-1}$ (centre) and $alpha_{+1}$ (right).}
    \label{fig:IteratedThetaDisplacement}
\end{figure}

With $\gamma=0.1$ and $\tau=0.01$, Figure \ref{fig:IteratedXDisplacement} presents a comparison between the theoretically predicted iterated displacement (red dots) and the numerically simulated displacement (blue dots). Similarly, comparisons for the variables $\vartheta(t)$ and $\alpha_\mp(t)$ are presented in Figure \ref{fig:IteratedThetaDisplacement}.

\section*{Conclusions}
This study investigates the dynamics of a Purcell three-link microswimmer equipped with passive elastic torsional coils at the joints. By controlling the spontaneous curvature, effectively manipulating the equilibrium angles between the links, we aim to understand and control the swimmer's motion.

The analysis begins by linearising the system around an equilibrium point, decoupling the equations for the angular variables (describing orientation and shape) from those for the translational variables (describing the swimmer's displacement). Applying small amplitude harmonic controls to the angular subsystem, we find that it asymptotically converges to a steady harmonic solution for sufficiently long times.

While this linear analysis provides valuable insights, it does not predict net advancement of the microswimmer. However, a subsequent weak nonlinear analysis reveals that a translation along the central link orientation emerges.

Furthermore, employing tools from geometric control theory, we prove that suitable small controls can steer the system between any two fixed configurations near the equilibrium point, establishing that the system is small time locally controllable. We also provide estimates for the microswimmer's displacement, both out of equilibrium (see equation \eqref{displacement}) and at the equilibrium point (see equation \eqref{eq: displacement in ye}), when using piecewise constant controls.

These theoretical predictions are supported by numerical simulations, which show good agreement between the dynamics obtained using small sinusoidal controls within the linearised system and those obtained using periodic piecewise constant controls within the geometric control framework.

In conclusion, we demonstrate that our prediction of the microswimmer's displacement, derived using periodic piecewise constant controls and geometric control theory, accurately approximates the displacement achieved with small continuous oscillating controls, particularly for short time periods and over longer timescales. This suggests that other directions of motion, generated by higher-order Lie brackets, can also be achieved by combining piecewise constant controls (as defined in equation \eqref{eq:u}) that generate the first-order bracket. Consequently, all directions of motion within the vicinity of the equilibrium point can be reached, at least over extended time periods, by employing suitable small oscillating controls that can be approximated by these piecewise constant controls.

This work represents a key step towards the development of a comprehensive model for an elastic microswimmer robot, providing foundational insights into their controllability and motion planning. Future research will focus on motion planning, trajectory tracking, and optimal control problems associated with this system.

\appendix
\section{Entries of the $\mathbf{\mathcal{R}}(t)$ matrix}
\label{R}
The non-zero entries of the symmetric Grand Resistance Matrix  $\mathbf{\mathcal{R}}(t)$ are the follows, denoting as $s=\sin$ and $c=\cos$:
%\begin{widetext}
\begin{equation*}
\begin{split}
    \label{R11}
  %   \mathbf{\mathcal{R}}_{11}=   
%\frac{(\eta -\xi ) (\cos (2 \alpha_{-1}+\vartheta )+\cos (2\alpha_{1}+\vartheta)+\cos (2 \vartheta)-3 (\eta +\xi )}{2 \xi },
    \mathbf{\mathcal{R}}_{11}=   
\frac{(1 -\nu ) (c (2 \alpha_{-1}+\vartheta )+c (2\alpha_{1}+\vartheta)+c (2 \vartheta))-3 (1 +\nu )}{2 \nu },
\end{split}
    \end{equation*}
\begin{equation*}
\begin{split}
    \label{R12}
    %\mathbf{\mathcal{R}}_{12} =
     % \frac{(\eta -\xi ) (\sin (2 (\alpha_{-1} +\vartheta ))+\sin (2( \alpha_{+1}+\vartheta ))+\sin (2 \vartheta ))}{2 \xi },
      \mathbf{\mathcal{R}}_{12} =
      \frac{(1 -\nu ) (s (2 (\alpha_{-1} +\vartheta ))+s (2( \alpha_{+1}+\vartheta ))+s (2 \vartheta ))}{2 \nu }
\end{split}
    \end{equation*}
%\end{multline}
\begin{equation*}
\begin{split}
%\begin{multline}
 \label{R13}
    %\mathbf{\mathcal{R}}_{13} =
    %\frac{(\xi -\eta ) \sin (2 \alpha_{-1} +\vartheta )-2 \eta  \sin (\alpha_{-1} +\vartheta )+(\eta -\xi ) \sin (2 \alpha_{+1}+\vartheta )+2 \eta  \sin (\alpha_{+1}+\vartheta )}{4 \xi },
    \mathbf{\mathcal{R}}_{13} =
    \frac{(\nu -1 ) s (2 \alpha_{-1} +\vartheta )-2 s (\alpha_{-1} +\vartheta )+(1 -\nu ) s (2 \alpha_{+1}+\vartheta )+2s (\alpha_{+1}+\vartheta )}{4 \nu },
\end{split}
    \end{equation*}
\begin{equation*}
 \label{R14}
    %\mathbf{\mathcal{R}}_{14} =
   %-\frac{\eta  \sin (\alpha_{-1} +\vartheta )}{2 \xi },
   \mathbf{\mathcal{R}}_{14} =
   -\frac{s (\alpha_{-1} +\vartheta )}{2 \nu},
\qquad 
    %\mathbf{\mathcal{R}}_{15} =
    %\frac{\eta  \sin (\alpha_{+1}+\vartheta )}{2 \xi }.
    \mathbf{\mathcal{R}}_{15} =
    \frac{s (\alpha_{+1}+\vartheta )}{2 \nu}.
\end{equation*}
\begin{equation*}
\begin{split}
 \label{R22}
    \mathbf{\mathcal{R}}_{22} =-\frac{(1 -\nu ) (c (2 (\alpha_{-1} +\vartheta ))+c (2 (\alpha_{+1}+\vartheta ))+c (2 \vartheta ))+3 (1 +\nu )}{2 \nu },
\end{split}
    \end{equation*}
\begin{equation*}
\begin{split}
 \label{R23}
    \mathbf{\mathcal{R}}_{23}=
    \frac{(1 -\nu ) c (2 \alpha_{-1} +\vartheta )+2 c (\alpha_{-1} +\vartheta )+(1 -\nu ) c (2 \alpha_{+1}+\vartheta )-2c (\alpha_{+1}+\vartheta )}{4 \nu},
\end{split}
    \end{equation*}
\begin{equation*}
    \label{R24}
     \mathbf{\mathcal{R}}_{24}=
  \frac{c (\alpha_{-1} +\vartheta )}{2 \nu},
\qquad
        \mathbf{\mathcal{R}}_{25}=    
-\frac{c (\alpha_{+1}+\vartheta )}{2 \nu }.
\end{equation*}
\begin{equation*}
\begin{split}
    \label{R33}
    \mathbf{\mathcal{R}}_{33}= 
\frac{c (2 \alpha_{-1} ) (1 -\nu )-4c (\alpha_{-1} )-2 (2c (\alpha_{+1})+4+\nu)+c (2 \alpha_{+1}) (\nu -1)}{8 \nu },   
\end{split}
    \end{equation*}
\begin{equation*}
    \label{R34}
    \mathbf{\mathcal{R}}_{34}=-\frac{3 c (\alpha_{-1} )+4}{12 \nu},
\qquad
    \mathbf{\mathcal{R}}_{35}=-\frac{3 c (\alpha_{+1})+4}{12 \nu}, \qquad
    \mathbf{\mathcal{R}}_{44}=-\frac{1}{3 \nu },
\qquad
    \mathbf{\mathcal{R}}_{55}=-\frac{1}{3\nu}.
\end{equation*}
%\end{widetext}

\section{Geometric Control Theory: Fundamentals}
\label{appendiceB}
Consider the nonlinear affine control system:
\begin{equation}
\label{drift_sys}
\dot{\mathbf{y}} = \qv_0(\mathbf{y}) + \sum_{j=1}^m u_j \qv_j(\mathbf{y}),
\end{equation}

\begin{Def}[Equilibrium Point]
An equilibrium point of the control system \eqref{drift_sys} is a pair $(\mathbf{y}_e, \mathbf{u}_e) \in \mathbb{R}^n \times \mathbb{R}^m$ such that $\qv_0(\mathbf{y}_e) + \sum_{j=1}^m (\mathbf{u}_e)_j \qv_j(\mathbf{y}_e) = 0.$
\end{Def}

\begin{Def}[Small-Time Local Controllability]
Let $(\mathbf{y}_e, \mathbf{u}_e)$ be an equilibrium point of the control system \eqref{drift_sys}. The system is said to be small-time locally controllable at $[\mathbf{y}_e, \mathbf{u}_e]$ if, for any $\varepsilon > 0$, there exists $\delta > 0$ such that, for every $\mathbf{y}_0 \in B_\delta(\mathbf{y}_e) = \{\mathbf{y} \in \mathbb{R}^n : |\mathbf{y} - \mathbf{y}_e| < \delta\}$ and every $\mathbf{y}_1 \in B_\delta(\mathbf{y}_e)$, there exist measurable control functions $\mathbf{u}: [0, \varepsilon] \to \mathbb{R}^m$ satisfying:
$
|\mathbf{u}(t) - \mathbf{u}_e| \leq \varepsilon, \quad \forall t \in [0, \varepsilon],
$
and steering the system from $\mathbf{y}_0$ to $\mathbf{y}_1$ in time $\varepsilon$.
\end{Def}

\begin{Def}[Lie Algebra Rank Condition]
\label{LARC}
The control system \eqref{drift_sys} satisfies the Lie algebra rank condition at the equilibrium point $(\mathbf{y}_e, \mathbf{u}_e)$ if
\[
\mathcal{A}(\mathbf{y}_e, \mathbf{u}_e) = \mathbb{R}^n,
\quad 
\text{where}
\quad
\mathcal{A}(\mathbf{y}_e, \mathbf{u}_e) = \{\gv(\mathbf{y}_e) \mid \gv \in \text{Lie}(\{\qv_0, \qv_1, \ldots, \qv_m\})\}.
\]
\end{Def}

% The following theorem provides a necessary condition for small-time local controllability:
% \begin{theorem}[Necessary Condition for STLC]
% \label{necessary_LARC}
% Assume that the control system \eqref{drift_sys} is small-time locally controllable at the equilibrium point $(\mathbf{y}_e, \mathbf{u}_e)$ and that the vector fields $\qv_i$ ($i = 0, 1, \ldots, m$) are real analytic. Then, the control system \eqref{drift_sys} satisfies the Lie algebra rank condition at $(\mathbf{y}_e, \mathbf{u}_e)$.
% \end{theorem}

\begin{Def}[Iterated Lie Brackets]
\label{IteratedLieBrackets}
Let $X$ and $Y$ be $C^\infty$ vector fields on $\mathbb{R}^n$. The iterated Lie bracket $ad^k_X Y$ is defined recursively as follows:
\begin{equation}
%\begin{aligned}
    ad^0_X Y = Y, \quad
    ad^{k+1}_X Y = [X, ad^k_X Y] \quad \text{for} \quad k \in \mathbb{N}.
%\end{aligned}
\end{equation}
We denote by $\text{Br}(\{\qv_0, \qv_1, \ldots, \qv_m\})$ the set of formal iterated Lie brackets of \\
$\qv_0, \qv_1, \ldots, \qv_m$. This is the smallest subset of $\text{Lie}(\{\qv_0, \qv_1, \ldots, \qv_m\})$ that is closed under the Lie bracket operation.
\end{Def}

Let $\mathbf{h}$ be a generic element of the Lie algebra generated by $\{\qv_0,\qv_1,\ldots,\qv_m\}$, and let $\mathbf{h}(\mathbf{y}_e)$ be its evaluation at the equilibrium point. Note that $\mathbf{h}$ can be expressed as a linear combination of Lie brackets of various orders, where the vector fields $\qv_j$ (with $j = 0, 1, \ldots, m$) are considered to be of order 1.

We define $\delta_j(\mathbf{h})$ as the number of times the vector field $\qv_j$ appears in the expression for $\mathbf{h}$. For instance, for $\mathbf{h} = \qv_1$, we have:
\begin{equation}
\delta_0(\mathbf{h}) = 0, \quad \delta_1(\mathbf{h}) = 1, \quad \text{and} \quad \delta_j(\mathbf{h}) = 0 \quad \text{for} \,\, j = 2, \ldots, m.
\end{equation}

Similarly, for $\mathbf{h} = [\qv_1, \qv_2]$, we have:
\begin{equation}
\label{exampleSussmann}
\delta_0(\mathbf{h}) = 0, \quad \delta_1(\mathbf{h}) = 1, \quad \delta_2(\mathbf{h}) = 1, \quad \text{and} \quad \delta_j(\mathbf{h}) = 0 \quad \text{for} \,\, j = 3, \ldots, m.
\end{equation}

Let $\mathfrak{S}_m$ be the symmetric group on $\{1, \ldots, m\}$. For each permutation $\pi \in \mathfrak{S}_m$, let $\tilde{\pi}$ denote the automorphism of $\text{Lie}(\{\qv_0, \ldots, \qv_m\})$ defined by $\tilde{\pi}(\qv_0) = \qv_0$ and $\tilde{\pi}(\qv_i) = \qv_{\pi(i)}$ for $i = 1, \ldots, m$.

For $\mathbf{h} \in \text{Br}(\{\qv_0, \qv_1, \ldots, \qv_m\})$, we define
\[
\sigma(\mathbf{h}) := \sum_{\pi \in \mathfrak{S}_m} \tilde{\pi}(\mathbf{h}) \in \text{Lie}(\{\qv_0, \ldots, \qv_m\}).
\]

For instance, when $m = 2$, we have
\[
\sigma(\mathbf{h}) = \mathbf{h} + \tilde{\pi}(\mathbf{h}) = [\qv_1, \qv_2] + [\qv_2, \qv_1].
\]

\begin{Def}[Sussmann Condition]
\label{Sussmann}
For $\Lambda \in [0, +\infty]$, the control system \eqref{drift_sys} satisfies the Sussmann condition $S(\Lambda)$ if:
\begin{enumerate}
    \item The Lie algebra rank condition (LARC) in Definition \ref{LARC} holds at the equilibrium $(\mathbf{y}_e, \mathbf{u}_e)$.
    \item For every $\mathbf{h} \in \text{Br}(\{\qv_0, \qv_1, \ldots, \qv_m\})$ with $\delta_0(\mathbf{h})$ odd and $\delta_i(\mathbf{h})$ even for $i = 1, \ldots, m$, $\sigma(\mathbf{h})(\mathbf{y}_e)$ lies in the span of the $\gv(\mathbf{y}_e)$'s, where the $\gv$'s are elements of $\text{Br}(\{\qv_0, \qv_1, \ldots, \qv_m\})$ and satisfy:
    \[
    \Lambda \delta_0(\gv) + \sum_{k=1}^m \delta_k(\gv) < \Lambda \delta_0(\mathbf{h}) + \sum_{k=1}^m \delta_k(\mathbf{h}),
    \]
    with the convention that, when $\Lambda = +\infty$, the inequality becomes $\delta_0(\gv) < \delta_0(\mathbf{h})$.
\end{enumerate}
\end{Def}
\begin{theorem}
\label{01}
If, for some $\Lambda \in [0, 1]$, the control system \eqref{drift_sys} satisfies the Sussmann condition, then it is small-time locally controllable.
\end{theorem}

\bibliographystyle{siamplain}
\bibliography{bibliography, sample}

\end{document}